\documentclass[prd,aps,floats]{revtex4}
\usepackage{amsmath,amssymb}
\usepackage{eucal}
\usepackage{graphicx}

\newcommand{\be}{\begin{equation}}
\newcommand{\ee}{\end{equation}}
\newcommand{\bea}{\begin{eqnarray}}
\newcommand{\eea}{\end{eqnarray}}
\newcommand{\vap}{\varepsilon}
\newcommand{\pary}{\partial_{y}}
\newcommand{\parp}{\partial}

\begin{document}

\title{Fixed-Connectivity Membranes}

\author{Mark Bowick}

\affiliation{Department of Physics, Syracuse University, Syracuse,
NY 13244-1130, USA.}

\begin{abstract}
The statistical mechanics of flexible surfaces with internal
elasticity and shape fluctuations is summarized. Phantom and
self-avoiding isotropic and anisotropic membranes are discussed,
with emphasis on the universal negative Poisson ratio common to
the low-temperature phase of phantom membranes and all strictly
self-avoiding membranes in the absence of attractive interactions.
The study of crystalline order on the frozen surface of spherical
membranes is also treated.
\end{abstract}
\maketitle

\section{Introduction}
\label{SECT__Intro}

The statistical mechanics of polymers, which are {\em
one-dimensional} chains or loops to a first approximation, has
proven to be a rich and fascinating field.\cite{deg,deja,deba} The
success of physical methods applied to polymers relies on
universality {--} many of the macroscopic length scale properties
of polymers are independent of microscopic details such as the
chemical identity of the monomers and their interaction
potential.\cite{KantorJer1}

Membranes are two-dimensional ($2D)$ generalizations of polymers.
The generalization of polymer statistical mechanics to membranes,
surfaces fluctuating in three dimensions, has proven to be very rich
because of the richer spectrum of shape and elastic deformations
available. In contrast to polymers, there are {\em distinct}
universality classes of membranes distinguished by the nature of
their short-range order. There are crystalline, fluid and hexatic
membrane analogues of the corresponding phases of strictly
two-dimensional systems (monolayers) where shape fluctuations are
frozen.\cite{NP:1987,DavidJer1,DRNlh}

\begin{figure}[ht]
\centering
\includegraphics[width=2in]{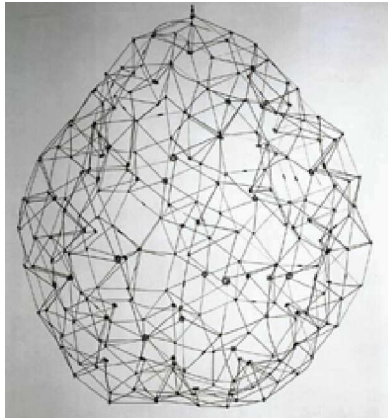}
\caption{Esfera (Sphere) 1976: Gertrude Goldschmidt (Gego).
Stainless steel wire {-} 97 $\times$ 88 cm (Patricia Phelps de
Cisneros Collection, Caracas, Venezuela).} \label{fig__esfera}
\end{figure}

The closest membrane analogue to a polymer is a $2D$ fishnet-like
mesh of nodes with a fixed coordination number for each node. A
fixed-connectivity membrane with spherical topology from the world
of art\cite{Storr:2003} is shown in Fig.\ref{fig__esfera}. Bonds
are assumed to be unbreakable while the nodes themselves live in
flat $d$-dimensional Euclidean space ${\bf R}^d$, with a physical
membrane corresponding to the case $d=3$. The intrinsic
crystalline order of fixed-connectivity membranes with, say,
typical coordination number 6, leads to the alternative
terminology crystalline membranes. They are also referred to as
polymerized or tethered membranes. In general the Hamiltonian for
a fixed-connectivity membrane will include both intrinsic elastic
contributions (compression and shear) and shape (bending)
contributions, since the membrane undergoes both types of
deformation.\cite{BT:2000,Wiese:2001}

Flexible membranes are an important member of the enormous class
of {\em soft} condensed matter
systems.\cite{deba,CL:1995,Lub:1997,Witten:1999} Soft matter
responds easily to external forces and has physical properties
that are often dominated by the entropy of thermal or other
statistical fluctuations.

This chapter will describe the properties of fixed-connectivity
membranes with focus on the universal negative Poisson ratio that
illustrates the novel elastic behavior of the extended ({\em
flat}) phase of physical membranes, the tubular phase of
anisotropic membranes and ordering on frozen curved membrane
topographies.

\section{Physical examples of membranes}
\label{SECT__Examples}

\begin{figure}[hb]
\centering
\includegraphics[width=2in]{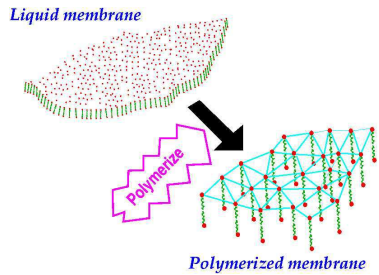}
\caption{Making a fixed-connectivity membrane by polymerizing a
fluid membrane.} \label{fig__polymerize}
\end{figure}

One can polymerize suitable chiral oligomeric precursors to form
molecular sheets.\cite{Stupp:1993} This approach is based directly
on the idea of creating an intrinsically two-dimensional polymer.
Alternatively one can permanently cross-link fluid-like
Langmuir-Blodgett films or amphiphilic layers by adding certain
functional groups to the hydrocarbon tails and/or the polar
heads,\cite{Fendler1:1984,Fendler2:1987} as shown schematically in
Fig.\ref{fig__polymerize}.

\begin{figure}[ht]
\centering
\includegraphics[width=2in]{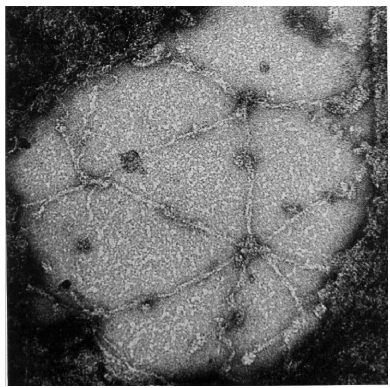}
\caption{An electron micrograph of a 0.5 micron square region of a
red blood cell cytoskeleton at magnification 365,000:1. The
skeleton is negatively stained and has been artificially spread to
a surface area nine to ten times as great as in the native
membrane. Image courtesy of Daniel Branton (Dept. of Biology,
Harvard University).} \label{fig__spectrin1}
\end{figure}

\begin{figure}[ht]
\centering
\includegraphics[width=2in]{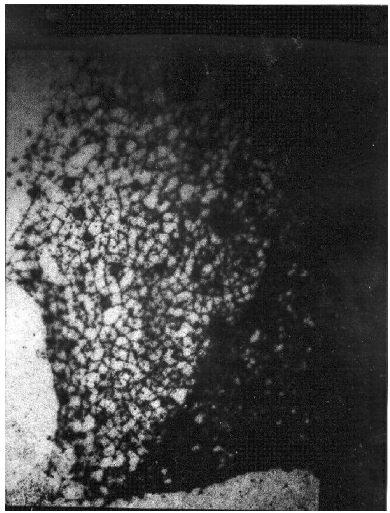}
\caption{An extended view of the spectrin/actin network which
forms the cytoskeleton of the red blood cell membrane. Image
courtesy of Daniel Branton.}
\label{fig__spectrin2}
\end{figure}

The 2D-cytoskeletons of certain cell membranes are beautiful and
naturally occurring fixed-connectivity membranes that are
essential to the function and stability of the cell as a
whole.\cite{Alberts:1994,Handbook:1995,PD:1999,Sackmann:2002,Boal:2002}
The simplest and most thoroughly studied example is the
cytoskeleton of mammalian erythrocytes (red blood cells). The
human body has roughly $5 \times 10^{13}$ red blood cells. The red
blood cell cytoskeleton is a protein network whose links are
spectrin tetramers (of length approximately 200 nm) meeting at
junctions composed of short actin filaments (of length 37 nm and
typically 13 actin monomers
long)\cite{ByBr:1985,ESMB:1986,Skel:1993} (see
Fig.\ref{fig__spectrin1} and Fig.\ref{fig__spectrin2}). There are
roughly 70,000 triangular faces in the entire mesh which is bound
as a whole by ankyrin and other proteins to the cytoplasmic side
of the other key component of the cell membrane, the fluid
phospholipid bilayer. Without the cytoskeleton the lipid bilayer
would disintegrate into a thousand little vesicles and certainly
the red blood cell would not be capable of the shape deformations
required to squeeze through narrow capillaries.

\begin{figure}[ht]
\centering
\includegraphics[width=2in]{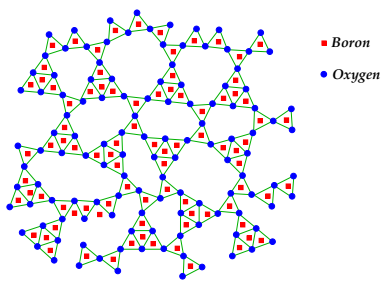}
\centering
\caption{The sheet molecule $B_2O_3$.}
\label{fig__sheet_mol}
\end{figure}

There are also inorganic realizations of fixed-connectivity
membranes. Graphitic oxide (GO) membranes are micron size sheets
of solid carbon, with thicknesses on the order of 10\AA, formed by
exfoliating carbon with a strong oxidizing agent. Their structure
in an aqueous suspension has been examined by several
groups.\cite{Hwa:1991,Wen:1992,Zasa:1994} Metal dichalcogenides
such as MoS$_2$ have also been observed to form rag-like
sheets.\cite{Chianelli:1979} Finally similar structures occur in
the large sheet molecules, shown in Fig.\ref{fig__sheet_mol}, and
believed to be an ingredient in glassy B$_2$O$_3$.

\section{Phase Diagrams}\label{SECT__POLYMEM}

\begin{figure}[ht]
\centering
\includegraphics[width=2in]{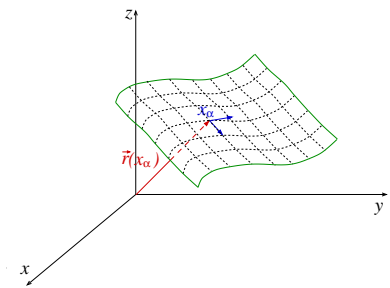}
\caption{The parametrization of a {\em membrane} with internal
coordinates ${\bf x}$ and bulk coordinates ${\vec r}({\bf x})$.}
\label{fig__quadr}
\end{figure}

Let us consider the general class of $D$-dimensional elastic and
flexible manifolds fluctuating in $d$-dimensional Euclidean space.
Such manifolds are described by a $d$-dimensional vector ${\vec
r}({\bf x})$, where ${\bf x}$ labels the $D$-dimensional internal
coordinates, as illustrated in Fig.\ref{fig__quadr}. A physical
membrane, of course, corresponds to the case $(D=2,d=3)$.

The Landau free energy of a membrane must be invariant under global
translations, so the order parameter is given by derivatives of the
embedding ${\vec r}$, viz. the tangent vectors ${\vec
t}_{\alpha}=\frac{\parp{\vec r}}{\parp x^{\alpha}}$, with
$\alpha=1,\cdots,D$. Invariance under rotations in both the internal
and bulk space limits the Landau free energy to the form
\cite{NP:1987,DRNJer1,PKN:88}

\bea\label{LAN_CR_VER} F({\vec r})&=&\int d^D{\bf
x}\,\left[\frac{t}{2}(\parp_{\alpha} {\vec r})^2 +
u(\parp_{\alpha} {\vec r}
\parp_{\beta} {\vec r} )^2 + v(\parp_{\alpha} {\vec r}
\parp^{\alpha} {\vec r})^2  +\frac{\kappa}{2} (\parp^2 {\vec
r})^2\right] \nonumber\\ &+& \frac{b}{2} \int d^D{\bf x}\, d^D{\bf
y}\, \delta^d \left({\vec r}({\bf x})- {\vec r}({\bf y})\right) \
, \eea

\noindent
where higher order terms are irrelevant in the long
wavelength limit. The physics of Eq.(\ref{LAN_CR_VER}) depends on
the elastic moduli $t$, $u$ and $v$, the bending rigidity $\kappa$
and the strength of self-avoidance $b$. The limit $b=0$ describes
a {\em phantom} membrane that may self-intersect with no energy
cost.

\begin{figure}[ht]
  \centering
  \includegraphics[width=4in]{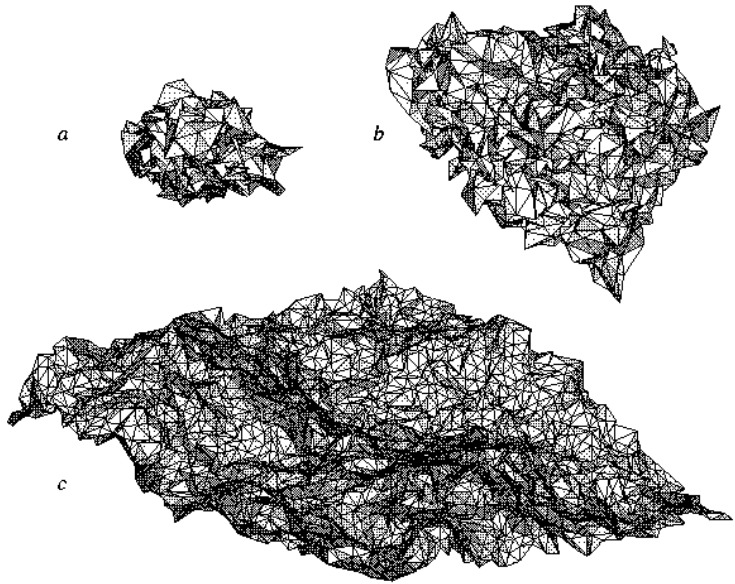}
  \caption{Typical configurations of phantom membranes: (a) the crumpled phase,
  (b) the critical {\em crumpling} phase and (c) the {\em flat} or
  bulk-orientationally-ordered phase. Images are from the simulations of
  Ref.31.}
  \label{fig__PHASES}
  \end{figure}

\noindent For small deformations from a reference ground state one
may write ${\vec r}({\bf x})$ as

\be\label{mf_variable} {\vec r}({\bf x})=\left(\zeta {\bf x}+{\bf
u}({\bf x}),{\vec h}({\bf x})\right) \ , \ee

\noindent where ${\bf u}({\bf x})$ are $D$ ``internal" phonon
modes and ${\vec h}({\bf x})$ $d-D$ ``out-of-plane" height
fluctuations. The case $\zeta=0$ corresponds to a mean field
isotropic crumpled phase for which typical equilibrium membrane
configurations have fractal Hausdorff dimension $d_H$ ($d_H =
\infty$ for phantom membranes) and there is no distinction between
the internal phonons and the height modes. The crumpled phase is
illustrated in Fig.\ref{fig__PHASES}(a).

The regime $\zeta \neq 0$ describes a membrane which is ``flat" up
to small fluctuations. The full rotational symmetry of the free
energy is spontaneously broken. The fields ${\vec h}$ are the
Goldstone modes and scale differently than the phonon fields ${\bf
u}$. Fig.\ref{fig__PHASES}(c) is a visualization of a typical
configuration in the ``flat" phase.

Phantom membranes are by far the easiest to treat analytically and
numerically. They may even be physically realizable by
synthesizing membranes from strands that cut and repair themselves
on a sufficiently short time scale that they access
self-intersecting configurations. One can also view the analysis
of the phantom membrane as the first step in understanding the
more physical self-avoiding membrane. Combined analytical and
numerical studies have yielded a thorough understanding of the
phase diagram of phantom fixed-connectivity membranes.

\subsection{Phantom Membranes}\label{SUBSECT__PHAN}

The phantom membrane free energy is

\be\label{LAN_CR_PH} F({\vec r})=\int d^D{\bf x} \,
\left[\frac{t}{2}(\parp_{\alpha} {\vec r})^2+u(\parp_{\alpha}
{\vec r}
\parp_{\beta} {\vec r} )^2+v(\parp_{\alpha} {\vec r}
\parp^{\alpha} {\vec r})^2 + \frac{\kappa}{2}(\parp^2
{\vec r})^2 \right] \ . \ee

\noindent The mean field effective potential, using the expansion
of Eq.(\ref{mf_variable}), is

\be\label{mf_eq_ph} V(\zeta)=D
\zeta^2\left(\frac{t}{2}+(u+vD)\zeta^2\right) \ , \ee

\noindent with minima

\be
\label{mf_zeta}
\zeta^2 = \left\{
             \begin{array}{r@{\quad:\quad}l}
              0 & t \ge 0 \\
              -\frac{t}{4(u+vD)} & t < 0 \ .
              \end{array}
              \right.
\ee

\begin{figure}[ht]
  \centering
  \includegraphics[width=3in]{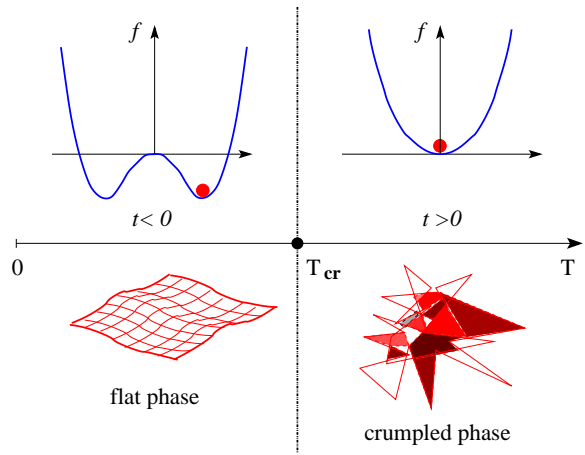}
  \caption{The mean field free energy density $f$ of fixed-connectivity
  membranes as a function of the order parameter $t$
  together with a schematic of the low temperature flat ordered phase and
  the high temperature crumpled disordered phase.}
  \label{fig__mfsoln}
  \end{figure}

\noindent This implies a ``flat" (extended) phase for $t < 0$ and
a crumpled phase for $t>0$, separated by a continuous crumpling
transition at $t=0$, as sketched in Fig.\ref{fig__mfsoln}.

\begin{figure}[ht]
  \centering
  \includegraphics[width=2.5in]{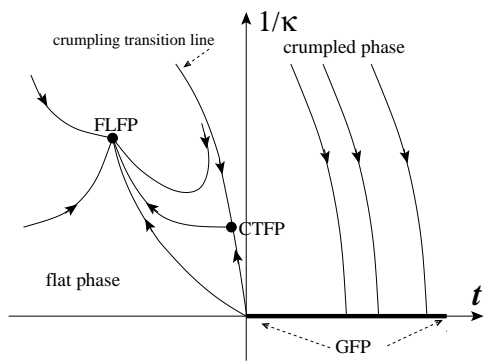}
  \caption{Schematic plot of the phase diagram for phantom
    membranes.}
  \label{fig__PHAN}
  \end{figure}

\noindent Of course anything is possible in mean field theory but
a variety of analytic and numerical calculations indicates the
true phase diagram of phantom membranes is qualitatively like
Fig.\ref{fig__PHAN}.\cite{BT:2000} The crumpled phase is described
by a line of equivalent Gaussian fixed points (GFPs).  There is a
crumpling transition line in the $\kappa-t$ plane containing an
infrared stable fixed point (CTFP) which describes the long
wavelength properties of the crumpling transition. Finally, for
large enough values of $\kappa$ and negative values of $t$, the
system is in a ``flat" phase whose properties are dictated by an
infrared stable flat phase fixed point (FLFP).

\subsubsection{The crumpled phase}\label{SUBSUBSECT__isocrphase}

In the crumpled phase, the free energy Eq.(\ref{LAN_CR_PH}), for
$D \ge 2$, simplifies to

\be
\label{LAN_CR_PH_IRR}
F({\vec r})=\frac{t}{2} \int d^D {\bf x}
\, (\parp_{\alpha} {\vec r})^2 + \mbox{irrelevant terms} \ ,
\ee

\noindent since the model is completely equivalent to a linear
sigma model with $O(d)$ internal symmetry in $D \ge 2$ dimensions
and therefore all derivative operators in ${\vec r}$ are
irrelevant by power counting.\cite{WK:1974,Goldenfeld:1992} The
parameter $t$ labels equivalent gaussian fixed points, as depicted
in Fig.\ref{fig__PHAN}. In renormalization group language there is
a marginal direction for positive $t$.  The large distance
properties of this phase are described by simple gaussian fixed
points with the exact connected 2-point function:

\be\label{CR_gg}
G({\bf x}) \sim \left\{
             \begin{array}{c  l}
              |{\bf x}|^{2-D}  \,  &:\, D\neq 2 \\
               \log\,|{\bf x}| \,  &:\,  D=2 \ .
              \end{array}
              \right.
\ee

\noindent The associated critical exponents may also be computed
exactly. The Hausdorff (fractal) dimension $d_H$, or equivalently
the size exponent $\nu=D/d_H$, is given (for the physical case
$D=2$) by

\be \label{CR_Haus} d_H=\infty \  (\nu = 0)  \Rightarrow R_G^2
\sim \log L \ , \ee

\noindent where $R_G^2$ is the radius of gyration and $L$ is the
linear membrane size. This result is confirmed by numerical
simulations of fixed-connectivity membranes in the crumpled phase
where the logarithmic behavior of the radius of gyration is
accurately
checked.\cite{BCFTA:1996}$^,$\cite{KKN:86}$^{-}$\cite{GK2:97}

\subsubsection{The Crumpling Transition}

\noindent
Near the crumpling transition the membrane free energy is
given by

\be
\label{LAN_CRTR_PH}
F({\vec r})=\int d^D{\bf x}\,\left[
\frac{1}{2}(\parp^2 {\vec r})^2+ u(\parp_{\alpha} {\vec r}
\parp_{\beta} {\vec r} )^2 + \hat{v}(\parp_{\alpha} {\vec r}
\parp^{\alpha} {\vec r})^2 \right] \ ,
\ee

\noindent where the bending rigidity has been scaled out and
$\hat{v} = v - \frac{u}{D}$. By naive power counting the
directions defined by the couplings $u$ and $\hat{v}$ are relevant
for $D \le 4$ and the model is amenable to an
$\vap=4-D$-expansion. The $\beta$ functions are given
by\cite{BT:2000}

\be \label{beta_FLAT_TR_1} \beta_u(u_R,v_R) = -\vap u_R +
\frac{1}{8\pi^2}\left\{\left(\frac{d}{3} +
\frac{65}{12}\right)u^2_R + 6u_R v_R + \frac{4}{3} v^2_R \right\}
\ee

\noindent and

\be \label{beta_FLAT_TR_2} \beta_v(u_R,v_R) =  -\vap v_R +
\frac{1}{8\pi^2}\left\{ \frac{21}{16}u^2_R + \frac{21}{2} u_R
v_R+(4d+5) v^2_R \right\} \ .
\ee

\noindent These two coupled beta functions have a fixed point only
for $d > 219$.\cite{DRNJer1} This suggests that the crumpling
transition is first order for $d=3$. Other analyses, however,
indicate a continuous crumpling transition.  A revealing extreme
limit of membranes was studied by David and Guitter.\cite{DG:88}
This is the limit of infinite elastic constants in the flat phase.
Since the elastic terms in the Hamiltonian scale like $q^2$ in
momentum space,  as compared to $q^4$ for the bending energy, this
limit exposes the dominant infrared behavior of the membrane. In
this ``stretchless" limit the elastic strain tensor
$u_{\alpha\beta}$ must vanish and the Hamiltonian is constrained,
very much in analogy to a nonlinear sigma model. The Hamiltonian
becomes

\be \label{CR_LGD} H_{NL}=\int d^D \sigma\, \frac{\kappa}{2}
(\Delta {\vec r})^2 \ ,
\ee

\noindent together with the constraint ${\parp_{\alpha} \vec r}
{\parp_{\beta} \vec r}= \delta_{\alpha \beta}$. Remarkably, the
$\beta$-function for the inverse bending rigidity
$\alpha=1/\kappa$ may be computed within a large-$d$ expansion,
giving

\be\label{dgbetafn} \beta(\alpha) = q\frac{\partial
\alpha}{\partial q} = \frac{2}{d}\,\alpha -\left(\frac{1}{4\pi} +
\frac{{\rm const}.}{d}\right)\alpha^2 \ . \ee

\noindent For $d=\infty$ there is no stable fixed point and the
membrane is always crumpled. To next order in $1/d$, however,
Eq.(\ref{dgbetafn}) reveals an ultraviolet stable fixed point at
$\alpha=8\pi/d$, corresponding to a continuous crumpling
transition. The size exponent at the transition is found to
be\cite{GDLP:88}

\be \label{nu_exp} d_H=\frac{2d}{d-1} \implies \nu=1-\frac{1}{d} =
2/3 \,\, (\mbox {for} \,\, d=3) \ . \ee

\noindent Le Doussal and Radzihovsky\cite{LDR:92} analyzed the
Schwinger-Dyson equations for the model of Eq.(\ref{LAN_CRTR_PH})
keeping up to four point vertices. The result for the Hausdorff
dimension and size exponent is

\be\label{CR_SCSA} d_H=2.73 \implies  \nu=0.73 \ . \ee

\noindent Finally Monte Carlo Renormalization Group
simulations\cite{ET:96} of the crumpling transition find a
continuous transition with exponents

\be\label{CR_MCRG} d_H=2.77(10) \implies \nu = 0.71(3) \ . \ee

\noindent Thus three independent analyses find a continuous
crumpling transition with a consistent size exponent.

  \begin{figure}[ht]
  \centering
  \includegraphics[width=3in]{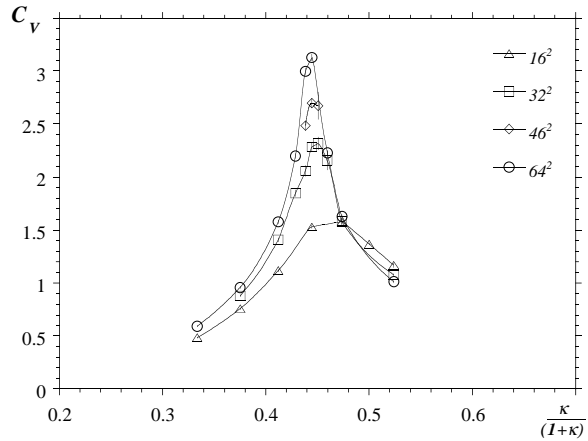}
  \caption{Plot of the specific heat observable from the
  simulations of Ref.31. The growth of the specific heat
  peak with system size indicates a continuous transition.}
  \label{fig__CR__cont}
  \end{figure}

\noindent Further evidence for the crumpling transition being
continuous is provided by numerous numerical
simulations\cite{BT:2000,GK1:97,GK2:97} where the analysis of
observables like the specific heat (see Fig.\ref{fig__CR__cont})
or the radius of gyration radius give textbook continuous phase
transitions, although the value of the exponents at the transition
are difficult to determine precisely.

\subsubsection{The Flat Phase}\label{SUB__subflat}

\begin{figure}[ht]
\centering
\includegraphics[width=2in]{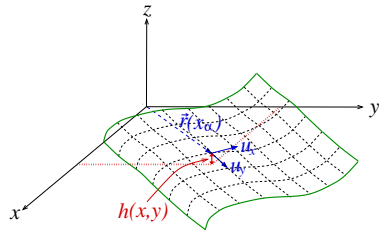}
\caption{Membrane coordinates appropriate for analyzing
fluctuations in the flat phase.} \label{fig__quadr_u}
\end{figure}

In a flat membrane (see Fig.\ref{fig__quadr_u}), it is natural to
introduce the strain tensor

\be \label{def_strain}
u_{\alpha \beta}= \partial_{\alpha}
u_{\beta}+\partial_{\beta} u_{\alpha} +\partial_{\alpha} h
\partial_{\beta} h \ .
\ee

\noindent
The free energy Eq.(\ref{LAN_CR_PH}) in these variables
becomes

\be\label{LAN_FL_PH}
F({\bf u},h)=\int d^D {\bf x} \left[
\frac{\hat{\kappa}}{2} (\parp_{\alpha} \parp_{\beta} h )^2+ \mu
u_{\alpha \beta} u^{\alpha \beta} + \frac{\lambda}{2}
(u^{\alpha}_{\alpha})^2 \right] \ ,
\ee

\noindent where irrelevant higher derivative terms have been
dropped. One recognizes the standard Landau free energy of
elasticity theory,\cite{Landau7} with Lam\'e coefficients $\mu$
and $\lambda$, plus an extrinsic curvature term, with bending
rigidity $\hat \kappa$. These couplings are related to the
original ones in Eq.(\ref{LAN_CR_PH}) by $\mu=u\zeta^{4-D}$,
$\lambda=2 v \zeta^{4-D}$, $\hat{\kappa}=\kappa \zeta^{4-D}$ and
$t=-4(\mu+\frac{D}{2}\lambda)\zeta^{D-2}$.

\begin{figure}[ht]
  \centering
  \includegraphics[width=2in]{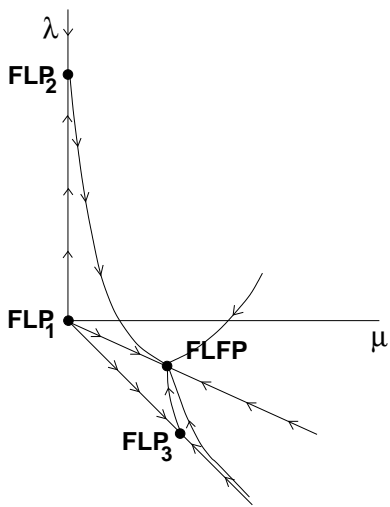}
  \caption{Phase diagram for the phantom flat phase. There are three
  infra-red unstable fixed points, labelled by FLP1, FLP2 and FLP3, but
  the physics of the flat phase is governed by the infra-red stable
  fixed point (FLFP).}
  \label{fig__PD_FLAT}
  \end{figure}

\noindent The large distance properties of the flat phase for
fixed-connectivity membranes are completely described by the free
energy of Eq.(\ref{LAN_FL_PH}). Since the bending rigidity may be
scaled out at the crumpling transition, the free energy becomes a
function of $\frac{\mu}{\kappa^2}$ and $\frac{\lambda}{\kappa^2}$.
The $\beta$-functions for the couplings $\mu$ and $\lambda$ in the
$\vap$-expansion are\cite{AL:1988,AGL:1989}

\bea\label{beta_CR_TR} \beta_{\mu}(\mu_R,\lambda_R)&=&-\vap \mu_R
      +\frac{\mu_R^2}{8 \pi^2}\left(\frac{d_c}{3}+20A\right) \ ;
      \\ \nonumber
\beta_{\lambda}(\mu_R,\lambda_R) &=& -\vap  \lambda_R+
      \frac{1}{8 \pi^2}\left(\frac{d_c}{3} \mu^2_R+2 (d_c+10A) \lambda_R \mu_R
      +2 d_c \lambda^2_R\right) \ ,
\eea

\noindent where $d_c$ is the codimension $d-D$ and
$A=\frac{\mu_R+\lambda_R}{2 \mu_R+\lambda_R}$. These coupled
$\beta$-functions possess four fixed points (see
Fig.\ref{fig__PD_FLAT}) whose values are shown in
Table~\ref{TAB__FL_EXP}.

The phase diagram revealed by the $\vap$-expansion is thus a
little more complex than that sketched in  Fig.\ref{fig__PHAN}.
The three additional fixed points are infra-red unstable, however,
and can only be reached for very specific values of the Lam\'e
coefficients.

  \begin{table}[hb]
  \caption{The fixed points and critical exponents of the flat
  phase.}
  \begin{center}
  \begin{tabular}{|c|c|c|c|c|}\hline
   FP      &    $\mu^\ast_R$ & $\lambda_R^\ast$ & $\eta$ & $\eta_u$\\\hline
   FLP1     &    $0$       &  $0$          & $0$  & $0$             \\\hline
   FLP2     &    $0$       &  $2\vap/d_c$  & $0$  & $0$             \\\hline
   FLP3     &    $\frac{12 \vap}{20+d_c}$  & $\frac{-6 \vap}{20+d_c}$
           &    $\frac{\vap}{2+d_c/10} $  & $\frac{\vap}{1+20/d_c}$ \\\hline
   FLFP    &    $\frac{12 \vap}{24+d_c}$  & $\frac{-4 \vap}{24+d_c}$
           &    $\frac{\vap}{2+d_c/12}$   & $\frac{\vap}{1+24/d_c}$ \\\hline
  \end{tabular}
  \end{center}
  \label{TAB__FL_EXP}
  \end{table}

\subsubsection{The properties of the flat phase}\label{SUBSUBSECT__flatphaseprops}

\noindent Fig.\ref{fig__PHASES}(c) shows a typical equilibrium
configuration for a membrane that has developed a preferred
orientation in the bulk {-} the surface normals clearly have
long-range order. In this phase the membrane is a rough extended
two-dimensional structure. The rotational symmetry of the full
free energy is spontaneously broken from $O(d)$ to $O(d-D) \times
O(D)$. The remnant rotational symmetry is realized in
Eq.(\ref{LAN_FL_PH}) as

\bea\label{sym__trans} h_i({\bf x}) &\rightarrow & h_i({\bf
x})+A^{i \alpha} {\bf x}_{\alpha} \ ;
\\\nonumber
u_{\alpha}({\bf x}) &\rightarrow& u_{\alpha} - A^{i \alpha} h_i
-\frac{1}{2} \delta^{i j} A^{i \alpha} A^{\beta j} {\bf x}_{\beta}
\ ,
\eea

\noindent where $A^{i \alpha}$ is a $D \times (d-D)$ matrix. This
relation provides Ward identities which greatly simplify the
renormalization of the theory.

\noindent
The phonon and height propagators in the infrared limit
are given by

\bea\label{low_q_mode} \Gamma_{u u}({\vec p}) &\sim & |{\vec
p}|^{2+\eta_u} \ ;
\\\nonumber
\Gamma_{h h}({\vec p}) &\equiv& |{\vec p}|^4 \kappa({\vec p}) \sim
|{\vec p}|^{4-\eta} \ , \eea where the last equation defines the
anomalous bending rigidity $\kappa({\vec p}) \sim |{\vec
p}|^{-\eta}$. The two scaling exponents $\eta_u$ and $\eta$ are
related by the scaling relation\cite{AL:1988}

\be\label{FLAT__scaling}
\eta_u=4-D-2\eta \ ,
\ee

\noindent which follows from the Ward identities
(Eq.(\ref{sym__trans})) associated with the remnant rotational
symmetry . The roughness exponent $\zeta$, which measures the
growth with system size of the rms height fluctuations transverse
to the flat directions, is determined from $\eta$ by the further
scaling relation

\be \zeta=\frac{4-D-\eta}{2} \ . \ee

\noindent The long wavelength properties of the flat phase are
described by the FLFP (see Fig.\ref{fig__PD_FLAT}). Since the FLFP
occurs at non-zero renormalized values of the Lam\'e coefficients,
the associated critical exponents are clearly non-Gaussian. These
key critical exponents have also been determined by independent
methods.

\noindent Large scale simulation of membranes in the flat phase
model were performed in Ref.\cite{BCFTA:1996} The results obtained
for the critical exponents are very accurate:

\be\label{Sim__exponents}
             \begin{array}{l l  l}
               \eta_u=0.50(1)\,; & \eta=0.750(5)\,; &
               \zeta=0.64(2)\, .
              \end{array}
\ee

\noindent A review of numerical results may be found in
Refs.\cite{GK1:97,GK2:97}.

\noindent The SCSA approximation\cite{LDR:92} gives a beautiful
result for general $d$: \be\label{SCSA_flat}
\eta(d)=\frac{4}{d_c+\sqrt{16-2d_c+d^2_c}} \ , \ee
which for $d=3$
gives \be\label{SCSA_flat_d3}
  \begin{array}{l l  l}
   \eta_u=0.358\,; & \eta=0.821\,; & \zeta=0.59 \ .
  \end{array}
\ee

\noindent Finally the large-d expansion\cite{DG:88} gives

\be\label{Larged_flat} \eta=\frac{2}{d} \implies \eta(3)=2/3 \ .
\ee

\noindent The numerical simulations are in qualitative agreement
with both the SCSA and large-d analytical estimates.

\noindent On the experimental side we are fortunate to have two
measurements of the key critical exponents for the flat phase of
fixed-connectivity membranes. The static structure factor of the
red blood cell cytoskeleton has been measured by small-angle x-ray
and light scattering, yielding a roughness exponent of
$\zeta=0.65(10)$.\cite{Skel:1993} Freeze-fracture electron
microscopy and static light scattering of the conformations of
graphitic oxide sheets reveal flat sheets with a fractal dimension
$d_H=2.15(6)$. Both these measured values are in good agreement
with the best analytic and numerical predictions, but the errors
are still too large to discriminate between different analytic
calculations and to accurately substantiate the numerical
simulations.

\noindent The Poisson ratio\cite{Landau7} of a phantom
fixed-connectivity membrane (which measures the transverse
elongation due to a longitudinal stress) is universal and within
the SCSA approximation is given by

\be\label{FL_PR} \sigma(D)=-\frac{1}{D+1} \implies \sigma(2)=-1/3
\ . \ee

\noindent This result has also been checked in numerical
simulations.\cite{ZDK:1996,FBGT:1997} Rather remarkably, it turns
out to be negative. While Ref.\cite{ZDK:1996} finds
$\sigma\approx-0.15$ the latter simulation\cite{FBGT:1997} finds
$\sigma\approx-0.32$. Materials with a negative Poisson ratio have
been dubbed {\em auxetics}\cite{ENHR:1991}. The wide variety of
potential applications of auxetic materials suggests a fascinating
role for flexible fixed-connectivity membranes in materials
science (see Sect.~\ref{SECT__PoissonRatio}).

A final critical regime of a flat membrane is achieved by
subjecting the membrane to external tension.\cite{GDLP:88} This
gives rise to a low temperature phase in which the membrane has a
domain structure, with distinct domains corresponding to flat
phases with different bulk orientations. This describes,
physically, a {\em buckled} membrane whose equilibrium shape is no
longer planar.

\subsection{Self-avoiding Membranes}\label{Sub_SECT__SA}

Physically realistic fixed-connectivity membranes will have large
energy barriers to self-intersection. That is they will generally
be self-avoiding. Self-avoidance is familiar in the physics of
polymer chains and may be treated by including the Edwards-type
delta-function repulsion of the Hamiltonian in
Eq.(\ref{LAN_CR_VER}). A detailed summary of our current
understanding is given in Refs.\cite{BT:2000,Wiese:2001} The
essential finding is that self-avoidance eliminates all but the
flat phase.

\subsubsection{Numerical simulations}

Numerical simulations are currently essential in understanding the
statistical mechanics of self-avoiding membranes because the
treatment of nonlinear elasticity together with non-local
self-avoidance is currently beyond the realm of analytic
techniques.

Two discretizations of membranes have been adopted to incorporate
self-avoidance. The {\em balls and springs} class of models begins
with a network of $N$ particles in a intrinsically triangular
array and interacting via a nearest-neighbor elastic potential

\be\label{tether_pot} V_{NN}({\vec r})=\left\{
\begin{array}{c c}
          0 & \mbox{for $|{\vec r}| \leq b$ } \\
          \infty & \mbox{for $|{\vec r}| > b $}
                         \end{array} \right.
                         \ ,
\ee

\noindent where the free parameter $b$ plays the role of a
tethering length. An additional hard sphere steric repulsion
forbids {\em any} node to be closer than a distance $\sigma$ from
any other node:

\be\label{exc_potential} V_{steric}({\vec r})=\left\{
\begin{array}{c c}
          \infty & \mbox{for $|{\vec r}| \leq \sigma $} \\
          0 & \mbox{ for $|{\vec r}| > \sigma $ }
                         \end{array} \right.
                         \ .
\ee

\noindent Early simulations\cite{KKN:86,KKN:87} of this class of
model gave a first estimate of the fractal dimension for physical
membranes compatible with the Flory estimate
$d_H=2(d+D)/(2+D)=2.5$.\cite{Duplantier} The system sizes
simulated, however, were quite small and subsequent simulations
for larger systems found that the membrane is
flat.\cite{PB:1988,ARP:1989} This result is remarkable when one
recalls that there is no explicit bending rigidity.

A plausible explanation\cite{AN1:90} for the loss of the crumpled
phase is that next-to-nearest neighbor excluded volume effects
induce a positive bending rigidity, driving the model to the FLFP.
The structure function of the self-avoiding model has been
computed numerically\cite{AN2:90} and found to compare well with
the analytical structure function for the flat phase of phantom
fixed-connectivity membranes. In particular the roughness
exponents are comparable.

The induced bending rigidity may be lowered by taking a smaller
excluded volume.\cite{BLLP:89} The flat phase persists to very
small values of $\sigma$ with eventual signs of a crumpled phase,
probably due to effective loss of self-avoidance. A more
comprehensive study,\cite{KK:93} in which the hard sphere radius
is taken to zero with an excluded volume potential which is a
function of the internal distance along the lattice, concluded
that self-avoidance implies flatness in the thermodynamic limit of
large membranes.

Self-avoidance may also be implemented by modelling impenetrable
triangular meshes. This has the advantage that there is no
restriction on the bending angle between adjacent cells and
therefore no induced bending rigidity.\cite{BCTT:2000}

The first simulations of the plaquette model\cite{B:91} found a
Hausdorff dimension in rough agreement in agreement with the Flory
estimate 2.5 but this has not held up in subsequent work. A
subsequent simulation\cite{KG:93} found $d_H \approx 2.3$ and
extensive recent work employing more sophisticated algorithms and
extending to much larger membranes confirm the loss of the
crumpled phase.\cite{BCTT:2000}

Some insight into the lack of a crumpled phase for self-avoiding
fixed-connectivity membranes is offered by the study of folding.
\cite{DiFGu1:94}$^{-}$\cite{CGP2:96} Folding corresponds to the
limit of infinite elastic constants\cite{DG:88} with the further
approximation that the space of bending angles is discretized. One
quickly discovers that the reflection symmetries of the allowed
folding vertices forbid local folding (crumpling) of surfaces.
There is therefore essentially no entropy for crumpling. There is,
however, local unfolding and the resulting statistical mechanical
models are non-trivial. The lack of local folding is the discrete
equivalent of the long-range curvature-curvature interactions that
stabilize the flat phase. The dual effect of the integrity of the
surface (time-independent connectivity) and self-avoidance is so
powerful that crumpling seems to be impossible in low embedding
dimensions.

\subsubsection{The properties of the self-avoiding fixed point}

\begin{figure}[ht]
\centering
\includegraphics[width=3in]{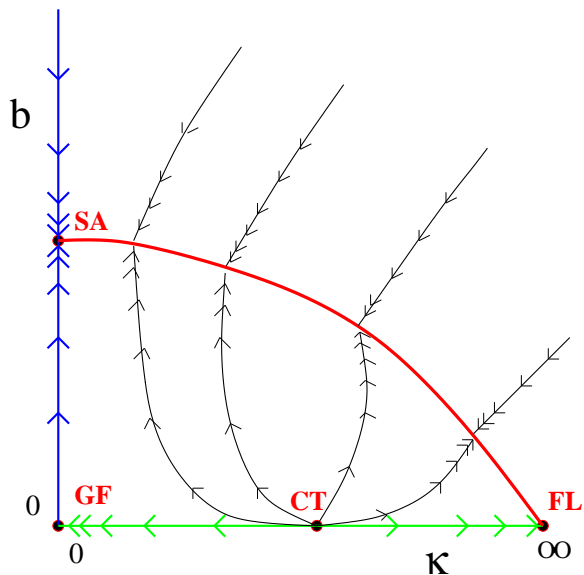}
\caption{The conjectured phase diagram for self-avoiding
fixed-connectivity membranes in 3 dimensions. With any degree of
self-avoidance the renormalization group flows are to the flat
phase fixed point of the phantom model (FL).}
\label{fig__SApd}
\end{figure}

For the physically relevant case $d=3$ numerical simulations thus
find that there is no crumpled phase. Furthermore, the flat phase
is {\em identical} to the flat phase of the phantom
membrane.\cite{BCTT:2000} The roughness exponent $\zeta_{SA}$ from
numerical simulations of self-avoidance at $d=3$ using
ball-and-spring models\cite{Grest:91} and impenetrable plaquette
models\cite{BCTT:2000} and the roughness exponent at the FLFP,
Eq.(\ref{Sim__exponents}), compare extremely well

\be \label{COMP_SA_FL} \zeta_{SA}=0.64(4) \ , \ \zeta=0.64(2) \
.
\ee

\noindent The numerical evidence thus strongly indicates that the
SAFP is exactly the same as the FLFP and that the crumpled
self-avoiding phase is absent in the presence of purely repulsive
potentials (see Fig.\ref{fig__SApd}). This conjecture is
strengthened by the finding that the Poisson ratio of
self-avoiding membranes is the same as that of flat phantom
membranes.\cite{BCTT:2001} (see Sect.\ref{SECT__PoissonRatio}).
This identification of fixed points enhances the significance of
the FLFP treated earlier.

\section{Poisson Ratio and Auxetics}
\label{SECT__PoissonRatio}

\noindent In the classical theory of elasticity\cite{Landau7} an
arbitrary deformation of a $D$-dimensional elastic body may be
decomposed into a pure shear and a pure compression:

\be\label{elasticdef} u_{ij}= [u_{ij} - \frac{1}{D}({\rm
Tr}\,u)\delta_{ij}] + \frac{1}{D}({\rm Tr}\,u)\delta_{ij} \ , \ee

\noindent
where Tr denotes the trace and the term in square
brackets is a pure shear (volume-preserving but shape changing)
while the second term is a pure compression (shape-preserving but
volume-changing). The elastic free energy is then given by
\be\label{elasticfe} F_{el} = \mu\left[u_{ij} - \frac{1}{D}({\rm
Tr}\,u)\delta_{ij}\right]^2 + \frac{1}{2}K\left({\rm
Tr}\,u\right)^2 \ , \ee

\noindent where $\mu$ is the shear modulus and $K$ is the bulk
modulus. This free energy may be written equivalently as

\be \label{Lame} F_{el} = \mu\,u_{ij}u_{ij} +
\frac{1}{2}\lambda\left({\rm Tr}\,u\right)^2 \ , \ee

\noindent
with the elastic Lam{\'e} coefficient $\lambda$ related
to the bulk and shear moduli by

\be\label{elmodrelns} K = \lambda + \frac{2\mu}{D} \ . \ee

\noindent For the physical membrane, $D=2$, this reads $K=\lambda
+ \mu$. Thermodynamic stability requires that both $K$ and $\mu$
be positive, otherwise the free energy could be spontaneously
lowered by pure compressional or pure shear deformations,
respectively.

The Poisson ratio $\sigma$ is defined as the ratio of transverse
contractile strain to longitudinal tensile strain for an elastic
body subject to a uniform applied tension T. For tension applied
uniformly in, say, the $x$-direction

\be
\label{PoissonRatio}
\sigma = - \frac{\delta y/y}{\delta x/x} \
, \ee

\noindent
the Poisson ratio is easily found to be

\be \sigma = \frac{K - \mu}{K + \mu} \ ,
\ee

\noindent for $D=2$, and

\be \sigma = \frac{1}{2} \left(\frac{3K - 2\mu}{3K + \mu}\right)
\ee

\noindent for $D=3$. Thermodynamic stability is only possible for
$-1 \leq \sigma \leq 1$ for $D=2$ and $-1 \leq \sigma \leq
\frac{1}{2} $ for $D=3$. The upper bounds (1 and 1/2 respectively)
are approached for materials that have vanishing shear modulus
compared to their bulk modulus (rubber-like) and the lower bounds
(-1) for materials with negligible bulk modulus in comparison to
their shear modulus (``anti-rubber")\cite{Poissonfootnote}.
Clearly, the Poisson ratio may be negative (auxetic) for $K < \mu$
(D=2) and $K < \frac{2}{3} \mu $ (D=3). Most materials get thinner
when stretched and fatter when squashed {--} auxetic materials are
uncommon. The earliest known example, dating from more than a
century ago, is that of a pyrite (FeS$_2$)
crystal\cite{Love:1944}, which has a Poisson ratio, in certain
crystallographic directions, of $\sigma \approx -0.14$. More
recently, some isotropic polyester foams have been created with
Poisson ratios as large as $\sigma \sim
-0.7$\cite{Lakes:1987,LakesURL}. The potential of auxetic
materials in materials science is nicely reviewed in
Ref.\cite{EvAl:2000}. One of the rare naturally occurring auxetics
is SiO$_2$ in its $\alpha${-}crystobalite
phase.\cite{YWP:1992,KC:1992} Cristobalite is one of the three
distinct crystalline forms of SiO$_2$, together with quartz and
tridymite. Its Poisson ratio reaches a maximum negative value of
$-0.5$ in some directions, with orientationally-averaged values
for single-phased aggregates of $-0.16$.

\begin{figure}[ht]
\centering
\includegraphics[width=3in]{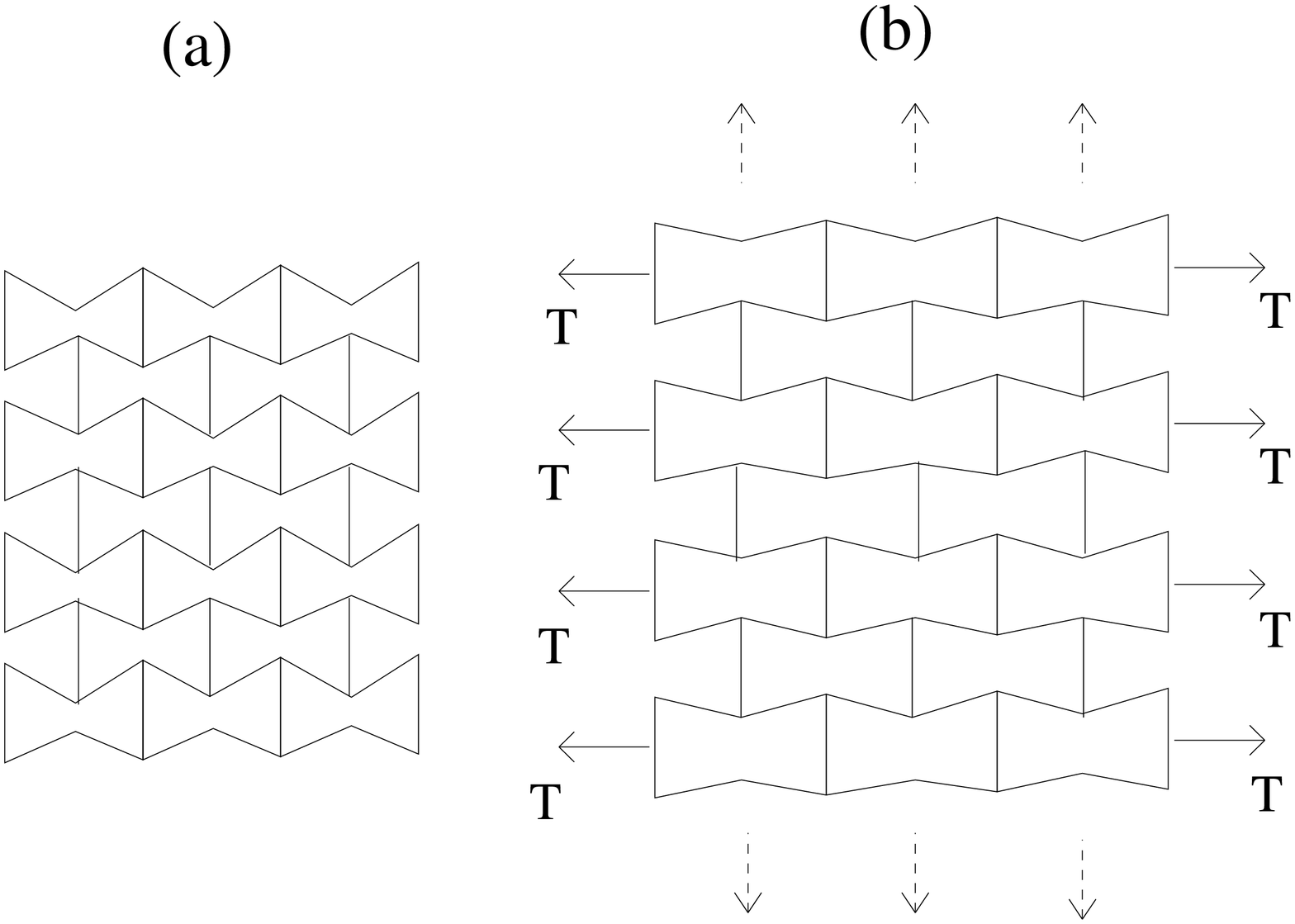}
\caption{Mechanical model of an auxetic material: (a) in the
absence of applied stress and (b) under applied lateral stress
$T$.  The lateral stretching accompanying the applied stress
forces the material out in the transverse dimension.}
\label{fig__auxmodel}
\end{figure}

The underlying mechanism driving fixed-connectivity membranes
auxetic ($\sigma=-1/3$) has schematic similarities to that
illustrated in Fig.\ref{fig__auxmodel}. Submitting a membrane to
tension will suppress its out-of-plane fluctuations, forcing it
entropically to expand in both in-plane directions. More
physically, the out-of-plane undulations renormalize the elastic
constants (the Lam\'e coefficients), in such a way that the
long-wavelength bulk modulus is less than the shear modulus, which
is the signature of a two-dimensional auxetic material. The soft
matter origin of the universal negative Poisson ratio of
fixed-connectivity membranes provides a fundamentally new paradigm
for the design of novel materials. The best current experimental
measurements of the Poisson ratio of the red blood cell
cytoskeleton\cite{DME:1994} find $\sigma\approx+1/3$ from separate
determinations of the bulk and shear modulus. The cytoskeleton
still has the fluid lipid bilayer attached, however, and this may
influence the pure cytoskeletal elasticity. A direct measurement
of the Poisson ratio for a flexible fixed-connectivity membrane
remains an important and challenging task.

Auxetic materials have desirable mechanical properties such as
higher in-plane indentation resistance, transverse shear modulus
and bending stiffness. They have clear applications as sealants,
gaskets and fasteners. They may also be promising materials for
artificial arteries, since they can expand to accommodate sudden
increases in blood flow.

We can model a realistic fixed-connectivity membrane with an
elastic free energy and either large bending rigidity or
self-avoidance. This is of practical importance in modelling
since, for example, we may replace the more complicated non-local
self-avoidance term with a large bending rigidity.

It would be very interesting to know if nature utilizes the
auxetic character of the red-blood cell spectrin cytoskeleton in
the elastic deformations of red blood cells as they pass through
fine blood capillaries. As such cells deform, the membrane
skeleton can unfold, which might help to transport large molecules
or expose reactive chemical groups.\cite{LakesNewsandViews:2001}

\section{Anisotropic Membranes}\label{SECT__POLYMEM_ANI}

An anisotropic membrane is a fixed-connectivity membrane in which
the elastic moduli or the bending rigidity in one distinguished
direction are different from those in the remaining $D-1$
directions. Such a membrane may be described by a $d$-dimensional
vector ${\vec r}({\bf x}_{\perp},y)$, where now the $D$
dimensional internal coordinates are split into $D-1 \ \ {\bf
x}_{\perp}$ coordinates and the orthogonal distinguished direction
$y$.

Requiring invariance under translations, $O(d)$ rotations in the
embedding space and $O(D-1)$ rotations in the internal space, the
equivalent of Eq.(\ref{LAN_CR_VER}) becomes

\bea \label{LGW} F\left(\vec r({\bf x})\right)&=& \frac{1}{2} \int
d^{D-1}{\bf x}_{\perp}\,dy \left[
\kappa_{\perp}(\partial_{\perp}^2 \vec r)^2 + \kappa_y (\pary^2
\vec r)^2 \right.
\nonumber\\
&& + \kappa_{\perp y} \pary^2 \vec r \cdot \parp_{\perp}^2 \vec r +
t_{\perp}(\parp_{\alpha}^{\perp} \vec r)^2 + t_y(\pary \vec r)^2
\nonumber\\
&& + \frac{u_{\perp \perp}}{2}(\parp_\alpha^{\perp} \vec r \cdot
\parp_{\beta}^{\perp} \vec r)^2 + \frac{u_{yy}}{2}(\pary \vec r \cdot
\pary \vec r)^2
\nonumber\\
&& + u_{\perp y} (\parp_{\alpha}^{\perp} \vec r \cdot \pary \vec r)^2
+ \frac{v_{\perp \perp}}{2}(\parp_{\alpha}^{\perp} \vec r \cdot
\parp_{\alpha}^{\perp} \vec r)^2
\nonumber\\
&& \left.  + v_{\perp y}(\parp_{\alpha}^{\perp} \vec r)^2 (\pary
\vec r)^2 \right]
\nonumber\\
&&+ \frac{b}{2} \int d^D {\bf x} \int d^D {\bf x}^\prime \delta^d
(\vec r({\bf x}) - \vec r({\bf x}^\prime))\ .
\eea

\noindent This model has eleven free parameters {--} three
distinct bending rigidities, $\kappa_{\perp},\kappa_y$ and
$\kappa_{\perp y}$, seven elastic moduli, $t_{\perp}, t_y,
u_{\perp \perp}, u_{yy}, u_{\perp y}, v_{\perp \perp}$ and $
v_{\perp y}$, and the strength of self-avoidance coupling $b$.

\begin{figure}
\centering
\includegraphics[width=4in]{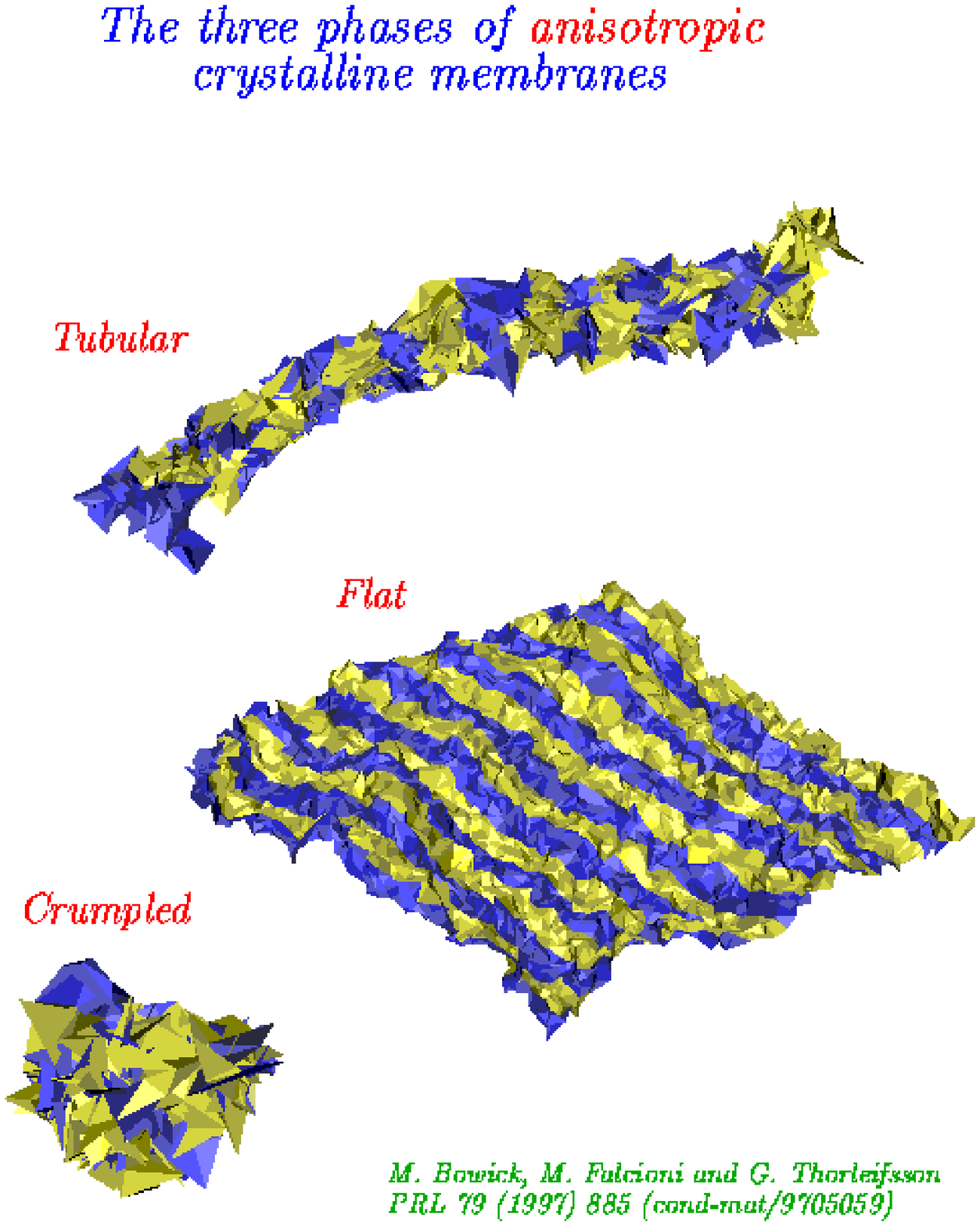}
\caption{} \label{fig__tubphases}
\end{figure}

As before we decompose displacements as

\be\label{mf_variable_any} {\vec r}({\bf x})=\left(\zeta_{\perp}
{\bf x}_{\perp}+{\bf u}_{\perp}({\bf x}),\, \zeta_y y +u_y({\bf
x}),\, {\vec h}({\bf x})\right) \ , \ee

\noindent with ${\bf u}_{\perp}$ being the $D-1${--}dimensional
intrinsic phonon modes, $u_y$ the intrinsic phonon mode in the
distinguished direction $y$ and ${\vec h}$ the $d-D$-dimensional
out-of-plane fluctuation mode. If $\zeta_{\perp}=\zeta_y=0$, the
membrane is crumpled and if both $\zeta_{\perp}$ and $\zeta_y$ do
not vanish the membrane is flat. There is, however, the
possibility that $\zeta_{\perp}=0$ and $\zeta_y \ne 0$ or
$\zeta_{\perp} \ne 0$ and $\zeta_y = 0$. This describes a {\em
tubular} phase, in which the membrane is crumpled in some internal
directions but flat in the remaining ones.\cite{RT:1995,RT:1998}.
Fig.\ref{fig__tubphases} displays a typical equilibrium
configuration from the tubular phase, along with the low and
high-temperature flat and crumpled phases for a phantom
anisotropic membrane.

Let's deal with the phantom anisotropic membrane first. Both
analytical\cite{R:2003} and numerical work\cite{BFT:1997} has
established that the phase diagram contains a crumpled, tubular
and flat phase. The crumpled and flat phases are equivalent to the
isotropic ones, so anisotropy turns out to be an irrelevant
interaction in those phases. The new physics is contained in the
tubular phase.

\subsection{Phantom Tubular Phase}\label{SECT__phase_ani}

\subsubsection{The Phase diagram}\label{SubSECT__tub}

We first describe the mean field theory phase diagram and then the
effect of fluctuations. There are two situations depending on the
value of a certain function $\Delta$, which depends on the elastic
constants $u_{\perp \perp},v_{\perp y},u_{yy}$ and $v_{\perp
\perp}$.\cite{RT:1995,RT:1998,R:2003}

\begin{figure}[ht]
  \centering
  \includegraphics[width=3in]{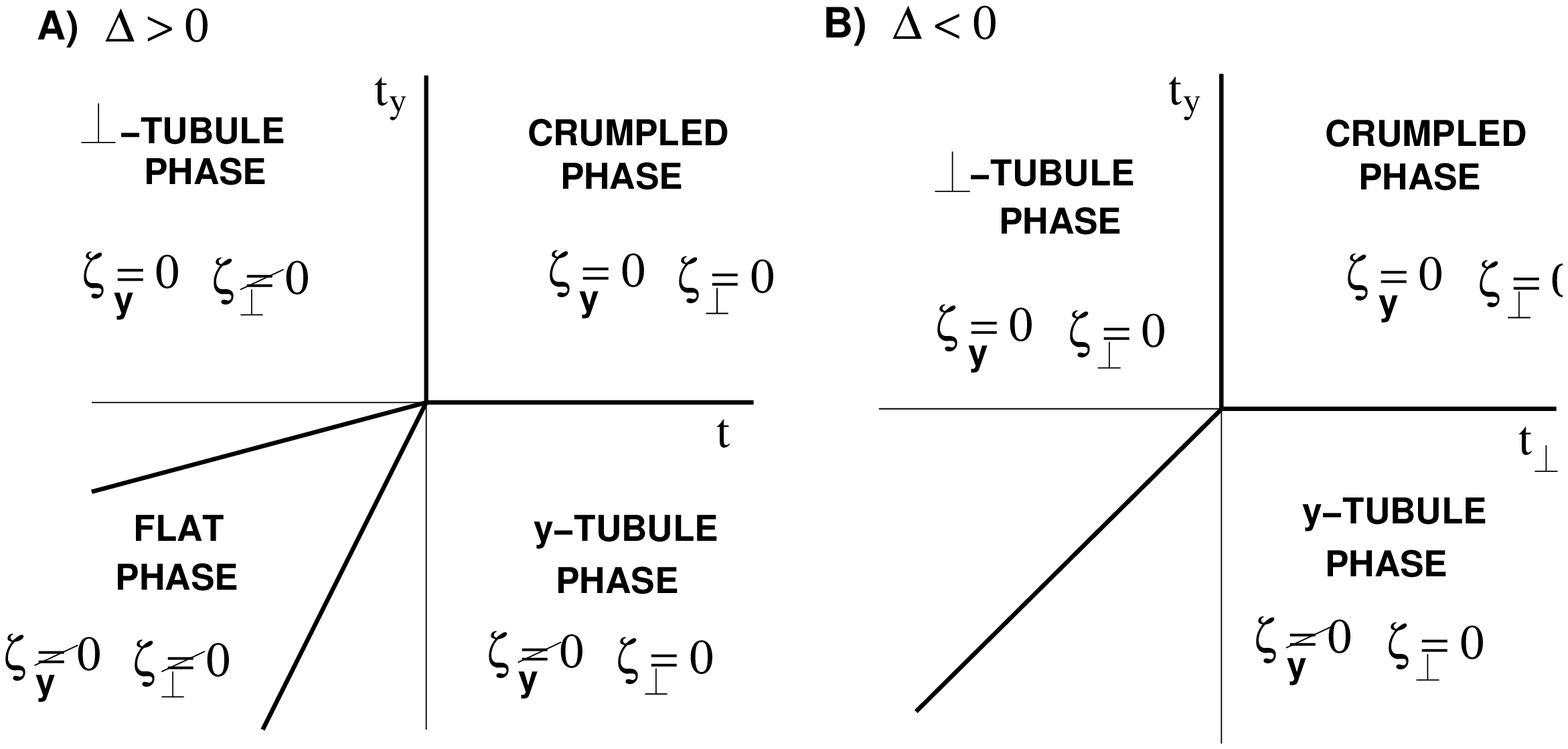}
  \caption{The phase diagram for anisotropic phantom membranes.}
  \label{fig__MF_an}
  \end{figure}

For $ \Delta > 0$ the mean field solution exhibits crumpled, flat
and tubular phases. When $t_y > 0$ and $t_{\perp}>0$ the model is
crumpled. Lowering the temperature so that one of the $t$
couplings becomes negative drives the membrane to the tubular
phase (either a $\perp$ or $y$-tubule). Lowering the temperature
still further flattens the membrane. For $ \Delta < 0$ the flat
phase disappears from the mean field solution, leaving only the
crumpled and tubular phases separated by a continuous transition.
Tubular phases are the stable low temperature stable phases in
this regime. This mean field result is summarized in
Fig.\ref{fig__MF_an}.

Beyond mean field theory, the Ginzburg criterion applied to this
particular model suggests that the phase diagram is stable for
physical membranes $D=2$ at any embedding dimension $d$. The mean
field description should be qualitative correct even for the full
model.

Numerical simulations have spectacularly confirmed this beautiful
analytic prediction.\cite{BFT:1997} Changing the temperature
generates a sequence of continuous phase transitions
crumpled-to-tubular and tubular-to-flat, in total agreement with
the $\Delta > 0$ case above (see Fig.\ref{fig__MF_an}).

\subsubsection{The Crumpled Anisotropic Phase}\label{SubSECT__cr_ani}

In this phase $t_y>0$ and $t_{\perp} >0$, and the free energy
Eq.(\ref{LGW}) reduces , for $D \ge 2$, to

\be\label{crum_ani_fe} F(\vec r({\bf x}))=\frac{1}{2} \int
d^{D-1}{\bf x}_{\perp} dy \left[ t_{\perp}(\parp_{\alpha}^{\perp}
\vec r)^2 + t_y(\pary \vec r)^2 \right]+\mbox{Irrelevant Terms} \
. \ee

\noindent By redefining the $y$ coordinate to be
$y^{\prime}=\frac{t_{\perp}}{t_y} y$ this reduces to
Eq.(\ref{LAN_CR_PH_IRR}), with $t\equiv t_{\perp}$. Anisotropy is
clearly irrelevant in the crumpled phase.

\subsubsection{The Flat Phase}\label{SubSECT__Fl_ani}

In the flat phase intrinsic anisotropies are only apparent at
short-distances and therefore should be irrelevant in the infrared
limit. This argument may be made more precise.\cite{Toner:88} The
flat phase is thus equivalent to the flat phase of isotropic
membranes.

\subsection{The Tubular Phase}\label{SECT__tubular}

We now turn to the study of the novel tubular phase, both in the
phantom case and with self-avoidance. Since the physically
relevant case for membranes is $D=2$ the $y$-tubular and
$\perp$-tubular phase are the same.

\begin{figure}[ht]
\centering
\includegraphics[width=3in]{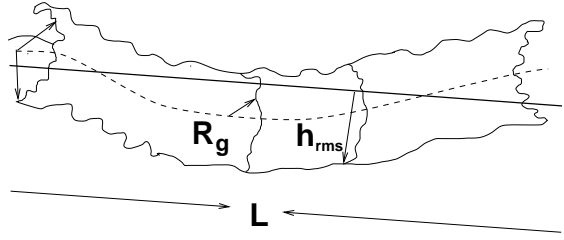}
\caption{A schematic illustration of a tubular configuration
indicating the radius of gyration $R_g$ and the height
fluctuations $h_{rms}$.} \label{fig__tubdef}
\end{figure}

The key critical exponents characterizing the tubular phase are
the size (or Flory) exponent $\nu$, giving the scaling of the
tubular diameter $R_g$ with the extended ($L_y$) and transverse
($L_{\perp}$) sizes of the membrane, and the roughness exponent
$\zeta$ associated with the growth of height fluctuations
$h_{rms}$ (see Fig.\ref{fig__tubdef}):

\bea
\label{nuzeta}
R_g(L_{\perp},L_y) & \propto & L_{\perp}^{\nu} S_R(L_y/L_{\perp}^z) \ ;\\
\nonumber h_{rms}(L_{\perp},L_y) & \propto & L_y^{\zeta}
S_h(L_y/L_{\perp}^z) \ ,
\eea

\noindent where $S_R$ and $S_h$ are scaling
functions\cite{RT:1995,RT:1998} and $z$ is the anisotropy
exponent.

\begin{figure}[ht]
   \centering
   \includegraphics[width=3in]{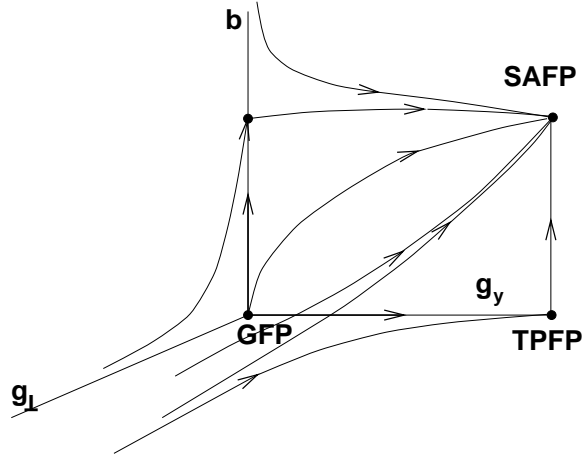}
   \caption{The phase diagram for self-avoiding anisotropic membranes
   with the Gaussian fixed point (GFP), the tubular phase fixed point
   (TPFP) and the self-avoidance fixed point (SAFP).}
   \label{fig__BG}
   \end{figure}

The general free energy described in Eq.(\ref{LGW}) may be
simplified considerably in a $y$-tubular phase.\cite{BG:97,BT:99}:

\bea\label{free_EG}
F(u,\vec h)&=&\frac{1}{2}\int d^{D-1}{\bf
x}_{\perp} dy \left[ \kappa (\pary^2 \vec h)^2+t(\parp_{\alpha}
\vec h)^2 \right.
\nonumber\\
&&+
g_{\perp}(\parp_{\alpha} u+\partial_{\alpha} \vec h \pary \vec h )^2
\nonumber\\
&&+\left.
g_y(\pary u+\frac{1}{2}(\pary \vec h)^2)^2
\right]
\nonumber\\
&&+ \frac{b}{2}\int dy d^{D-1}{\bf x}_{\perp}d^{D-1}{\bf
x}_{\perp}^\prime \delta^{d-1}(\vec{h}({\bf
x}_{\perp},y)-\vec{h}({\bf x}_{\perp}^\prime,y)) \ ,
\eea

\noindent reducing the number of free couplings to five. The
coupling $g_{\perp}$, furthermore, is irrelevant by standard power
counting. The most natural assumption is to set it to zero. In
that case the phase diagram one obtains is shown in
Fig.\ref{fig__BG}. Without self-avoidance, i.e. $b=0$, the
Gaussian Fixed Point (GFP) is unstable and the long-wavelength
behavior of the membrane is controlled by the tubular phase fixed
point (TPFP). Any amount of self-avoidance, however, leads to a
new fixed point, the Self-avoiding Tubular fixed point (SAFP),
which describes the large distance properties of self-avoiding
tubules.

\begin{figure}[ht]
\centering
\includegraphics[width=3in]{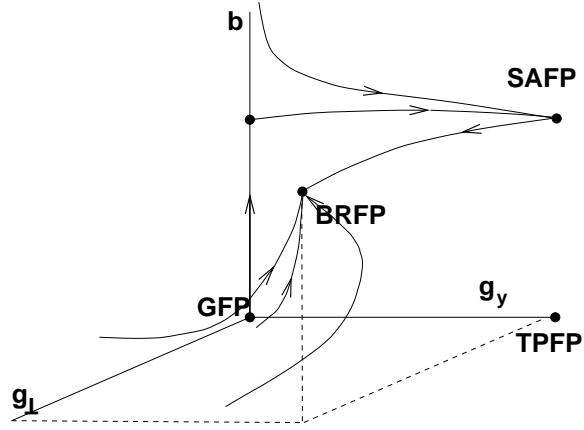}
\caption{The phase diagram for self-avoiding anisotropic membranes
with the Gaussian fixed point (GFP), the tubular phase fixed point
(TPFP), the self-avoidance fixed point (SAFP) and the bending
rigidity fixed point (BRFP).} \label{fig__RT}
\end{figure}

Radzihovsky and Toner advocate a different scenario.\cite{RT:1998}
For sufficiently small embedding dimensions $d$, including the
physical $d=3$ case, these authors suggest the existence of a new
bending rigidity renormalized fixed point (BRFP), which is the
infra-red fixed point describing the actual properties of
self-avoiding tubules (see Fig.~\ref{fig__RT}).

Here we follow the arguments presented in Refs.\cite{BG:97,BT:99}
and consider the model defined by Eq.(\ref{free_EG}) with the
$g_{\perp}$-term set to zero. One can prove then than there are
some general scaling relations among the critical exponents. All
three exponents may be expressed in terms of a single exponent

\bea\label{Ani_scaling}
\zeta&=&\frac{3}{2}+\frac{1-D}{2z} \ ;\nonumber\\
&\nu&=\zeta z \ .
\eea

Remarkably, the phantom case, as described by Eq.(\ref{free_EG}),
can be solved exactly. The result for the size exponent is

\be\label{Phantom_ani_size} \nu_{ph}(D)=\frac{5-2D}{4} \ , \,
\nu_{ph}(2)= \frac{1}{4} \ee with the remaining exponents
following from the scaling relations Eq.(\ref{Ani_scaling}).

The self-avoiding case may be treated with techniques similar to
those in isotropic case. The size exponent may be estimated within
the Flory approximation, yielding

\be\label{Flory_ani} \nu_{Fl}=2/d_H=\frac{D+1}{d+1} \ . \ee

\noindent The Flory estimate is an uncontrolled approximation.
Fortunately, a $\vap$-expansion, adapting the multi-local operator
product expansion technique\cite{DHK:1993}$^{-}$\cite{DDG:1997} to
the case of tubules, is also possible.\cite{BG:97,BT:99} The
resulting renormalization group $\beta$-functions provide evidence
for the phase diagram shown in Fig.\ref{fig__BG}. Extrapolation
techniques also provide estimates for the size exponent, the most
accurate value being $\nu=0.62$ for the physical case. The rest of
the exponents may be computed from the scaling relations.

Numerical simulations so far, however, do {\em not} find a tubular
phase in the case of strict self-avoidance.

\section{Order on Curved Surfaces}

Imagine we instantaneously freeze a fluctuating membrane so that
it has some fixed but curved shape. We can then ask about the
nature of the ground state of particles distributed on this
surface and interacting with some microscopic pair-wise repulsive
potential. The relevant physics is clearly related to the infinite
bending rigidity limit (flat phase) of elastic membranes. In such
a membrane the topology and topography are fixed.

\begin{figure}[ht]
\centering
\includegraphics[width=2in]{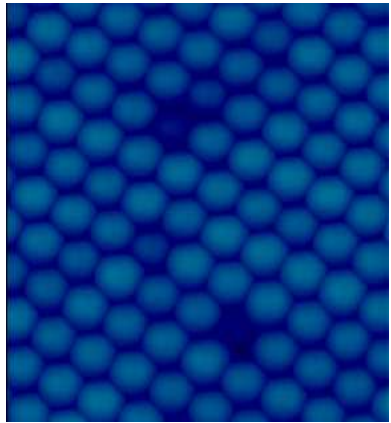}
\caption{A 2.5 micron scan of 0.269 micron diameter polystyrene
spheres crystallized into a regular triangular lattice - taken
from http://invsee.asu.edu/nmodules/spheresmod/.}
\label{fig__hexpack}
\end{figure}

Spherical particles on a flat surface pack most efficiently in a
simple triangular lattice, as illustrated in
Fig.\ref{fig__hexpack}. In the dense limit each particle ``kisses"
six of its neighbors.\cite{kissingno} Such six-coordinated
triangular lattices cannot, however, be perfectly wrapped on the
curved surface of a sphere; topology alone requires there be
defects in coordination number.  The panels on a soccer ball and
the spherical carbon molecule C$_{60}$
(buckyball)\cite{KHOCS:1985,J:2000} are good illustrations of the
necessity of defects for a spherical triangulation {--} they have
12 pentagonal faces (each the dual of a 5-coordinated defect) in
addition to 20 hexagonal faces (each dual to a regular
6-coordinated node.) The necessary packing defects can be
characterized by their topological charge, $q$, which is the
departure of their coordination number $c$ from the preferred flat
space value of 6 ($q=6-c$). These coordination number defects are
point-like topological defects called {\em
disclinations}\cite{David} and they detect intrinsic Gaussian
curvature located at the defect. A profound theorem of
Euler\cite{Euler:1750,HP:1996} states that the total disclination
charge of {\em any} triangulation whatsoever of the sphere must be
12!\cite{Sphere:12} A total disclination charge of 12 can be
achieved in many ways, however, which makes the determination of
the minimum energy configuration of repulsive particles, essential
for crystallography on a sphere, an extremely difficult problem.
This was recognized nearly 100 years ago by J.J.
Thomson\cite{JJT:1904}, who attempted, unsuccessfully, to explain
the periodic table in terms of rigid electron shells. Similar
problems arise in fields as diverse as multi-electron bubbles in
superfluid helium,\cite{Leiderer:1995} virus
morphology,\cite{CasparKlug:1962}$^{-}$\cite{LMN:2003} protein
s-layers,\cite{Sleytr:2001,PMS:1991} giant
molecules\cite{Liu:2002,MKD:2001} and information
processing.\cite{Sloane:1984,CS:1998} Indeed, both the classic
Thomson problem, which deals with particles interacting through
the Coulomb potential, and its generalization to other interaction
potentials, are still open problems after almost 100 years of
attention.\cite{Smale:1998}$^{-}$ \cite{EH:1997}

The spatial curvature encountered in curved geometries adds a
fundamentally new ingredient to crystallography not found in the
study of order in spatially flat systems. As the number of
particles on the sphere grows, isolated charge 1 defects (5s) will
induce too much strain. This strain can be relieved by introducing
additional dislocations, consisting of pairs of tightly bound 5-7
defects\cite{Cotterill:1985,Nelson:2002}, which don't spoil the
topological constraints because their net disclination charge is
zero. Dislocations, which are themselves point-like topological
defects in two dimensions, disrupt the translational order of the
crystalline phase but are less disruptive of orientational
order.\cite{Nelson:2002}

Recent work on an experimental realization of the generalized
Thomson problem has allowed us to explore the lowest energy
configuration of the dense packing of repulsive particles on a
spherical surface and to confront a previously developed theory
with experiment.\cite{Scars:2003}. We create two-dimensional
packings of colloidal particles on the surface of spherical water
droplets and view the structures with optical microscopy. Above a
critical system size, the thermally equilibrated colloidal
crystals display distinctive high-angle grain boundaries, which we
call ``scars". These grain boundaries are found to end entirely
within the crystal, which is never observed on flat surfaces
because the energy penalty is too high.

The experimental system is based on the self-assembly of one
micron diameter cross-linked polystyrene beads adsorbed on the
surface of spherical water droplets (of radius $R$), themselves
suspended in a density-matched oil mixture.\cite{DHNMBW:2002} The
polystyrene beads are almost equally happy to be in oil or water
(the bead/oil surface tension is close to the bead/water surface
tension) and therefore diffuse freely {\em until} they find the
oil-water interface and stick there. Particle assembly on the
interface of two distinct liquids dates to the pioneering work of
Pickering\cite{Pickering:1907} and was beautifully exploited by
Pieranski\cite{P:1980} some time ago. The particles are imaged
with phase contrast using an inverted microscope. After
determining the center of mass of each bead, the lattice geometry
is analyzed by original Delaunay triangulation
algorithms\cite{Delaunay:1934} appropriate to spherical surfaces.

\begin{figure}[ht]
\centering
\includegraphics[width=3in]{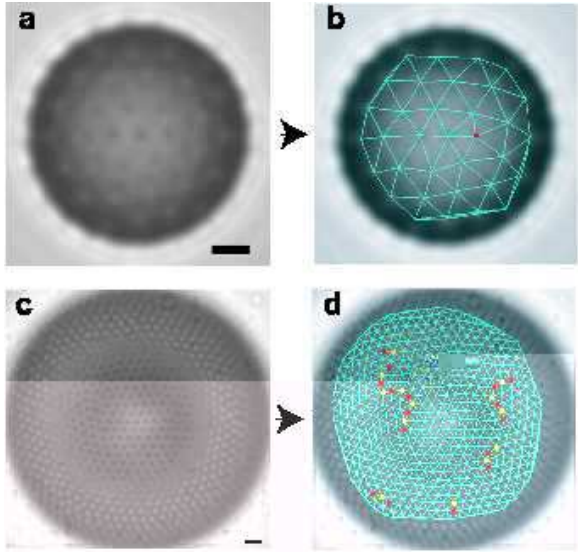}
\caption{Light microscope images of particle-coated droplets. Two
droplets ({\bf A}) and ({\bf C}) are shown, together with their
associated defect structures ({\bf B}) and ({\bf D}). Panel ({\bf
A}) shows an $\approx 13\%$ portion of a small spherical droplet
with radius $R = 12.0$ microns and mean particle spacing $a = 2.9$
microns (${\rm R/a} = 4.2$), along with the associated
triangulation ({\bf B}). Charge $+1(-1)$ disclinations are shown
in red and yellow respectively. Only one $+1$ disclination is
seen. Panel ({\bf C}) shows a cap of spherical colloidal crystal
on a water droplet of radius $R = 43.9$ microns with mean particle
spacing $a=3.1$ microns (${\rm R/a}=14.3$), along with the
associated triangulation ({\bf D}). In this case the imaged
crystal covers about $17\%$ of the surface area of the sphere. The
scale bars in ({\bf A}) and ({\bf C}) are 5 microns.}
\label{panel}
\end{figure}

We analyze the lattice configurations of a collection of 40
droplets.  A typical small spherical droplet with system size,
${\rm R/a}=4.2$, where $a$ is the mean particle spacing, is shown
in Fig.~\ref{panel}{\bf A}. The associated Delaunay triangulation
is shown in Fig.~\ref{panel}{\bf B}. The only defect is one
isolated charge $+1$ disclination. Extrapolation to the entire
surface of the sphere is statistically consistent with the
required $12$ total disclinations.

Qualitatively different results are observed for larger droplet
sizes as defect configurations with excess dislocations appear.
Although some of these excess dislocations are isolated, most
occur in the form of distinctive ($5-7-5-7-{\cdots}-5$) chains,
each of net charge $+1$, as shown in Fig.~\ref{panel}{\bf D}.
These chains form high-angle ($30^{\circ}$) grain boundaries, or
scars, which terminate freely within the crystal.  Such a feature
is energetically prohibitive in equilibrium crystals in flat
space. Thus, although grain boundaries are a common feature of 2D
and 3D crystalline materials, arising from a mismatch of
crystallographic orientations across a boundary, they usually
terminate at the boundary of the sample in flat space because of
the excessive strain energy associated with isolated terminal
disclinations. Termination within the crystal is a feature unique
to curved space.

\begin{figure}
\centering
\includegraphics[width=3.5in]{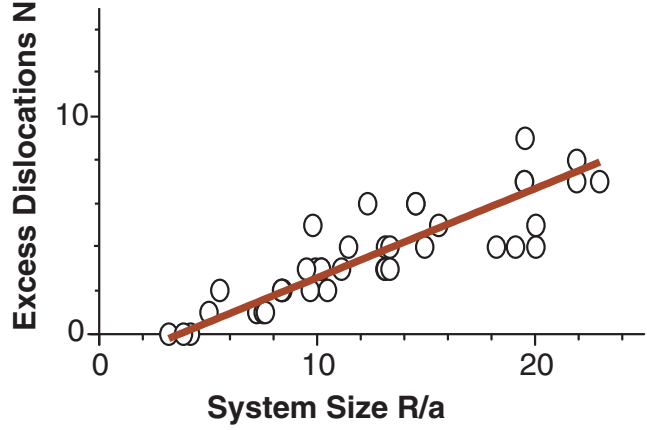}
\caption{Excess dislocations as a function of system size. The
number of excess dislocations per minimal disclination $N$ as a
function of system size $R/a$, with the linear prediction given by
theory shown as a solid red line.}
\label{NvsR}
\end{figure}

Of key interest is the number of excess dislocations per chain as a
function of the dimensionless system size ${\rm R/a}$. This is
plotted in Fig.~\ref{NvsR}. Scars only appear for droplets with
${\rm R/a} \geq 5$. These results provide a critical confirmation of
a theoretical prediction that ${\rm R/a}$ must exceed a threshold
value $({\rm R/a})_c \approx 5$, corresponding to $M \approx 360$
particles, for excess defects to proliferate in the ground state of
a spherical crystal.\cite{BNT:2000} The precise value of $({\rm
R/a})_c$ depends on details of the microscopic potential, but its
origin is easily understood by considering just one of the 12 charge
$+1$ disclinations required by the topology of the sphere. In flat
space such a topological defect has an associated energy that grows
quadratically with the size of the system,\cite{Nelson:2002} since
it is created by excising a $2\pi/6$ wedge of material and gluing
the boundaries together.\cite{CL:1995,Nelson:2002} The elastic
strain energy associated with this defect grows as the area. In the
case of the sphere the radius plays the role of the system size. As
the radius increases, isolated disclinations become much more
energetically costly. This elastic strain energy may be reduced by
the formation of linear dislocation arrays, i.e. grain boundaries.
The energy needed to create these additional dislocation arrays is
proportional to a dislocation core energy $E_c$ and scales linearly
with the system size.\cite{BNT:2000} Such screening is inevitable in
flat space (the plane) if one forces an extra disclination into the
defect-free ground state. Unlike the situation in flat space, grain
boundaries on the sphere can freely
terminate,\cite{BNT:2000}$^{-}$\cite{BCNT:2002} consistent with the
scars seen on colloidal droplets.

\begin{figure}
\centering
\includegraphics[width=2.5in]{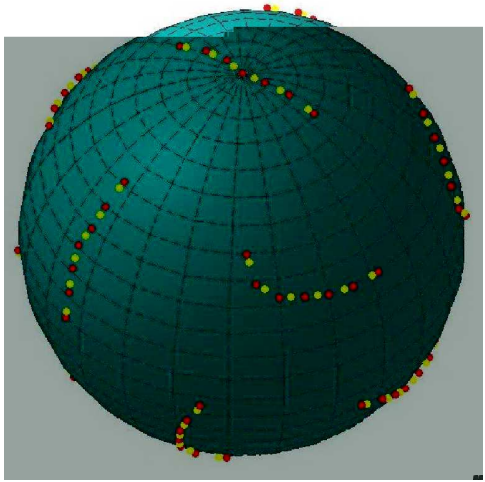}
\caption{Model grain boundaries. This image is obtained from a
numerical minimization for a system size comparable to the large
droplet in Fig.\ref{panel}({\bf c},{\bf d}).}
\label{theoryprediction}
\end{figure}

A powerful analytic approach to determining the ground state of
particles distributed on a curved surface has been
developed.\cite{BNT:2000,BCNT:2002,BNT:2003} The original particle
problem is mapped to a system of interacting disclination defects
in a continuum elastic curved background. The defect-defect
interaction is universal with the particle microscopic potential
determining two free parameters {--} the Young modulus $K_0$ of
the elastic background and the core energy $E_c$ of an elementary
disclination. A rigorous geometrical derivation of the effective
free energy for the defects is given in Ref.\cite{BT:2001} An
equivalent derivation may also be given by integrating out the
phonon degrees of freedom from the elastic
Hamiltonian,\cite{DRNJer1} with the appropriate modifications for
a general distribution of defects. The energy of a two-dimensional
crystal embedded in an arbitrary frozen geometry described by a
metric $g_{ij}({\bf x})$ is given by

\begin{eqnarray}
\label{inv_Lap}
H=&E_0& + \frac{Y}{2}\int\int d\sigma(x) d\sigma(y) \nonumber \\
 & \times & \left\{\left[s(x)-K(x)\right]\frac{1}{\Delta^2}\left[(s(y)-K(y)\right] \right\}
 \ ,
\end{eqnarray}

\noindent where the integration is over a fixed surface with area
element $d\sigma(x)$ and metric $g_{ij}$, $K$ is the Gaussian
curvature, $Y$ is the Young modulus in flat space and
$s(x)=\sum_{i=1}^N \frac{\pi}{3}q_i \delta(x,x_i)$ is the
disclination density $\left[\delta(x,x_i)=\delta(x-x_i)/\sqrt{det
(g_{ij})}\right]$. Here 5- and  7-fold defects correspond to
$q_i=+1$ and $-1$, respectively. Defects like dislocations or
grain boundaries can be built from these $N$ elementary
disclinations. $E_0$ is the energy corresponding to a perfect
defect-free crystal with no Gaussian curvature; $E_0$ would be the
ground state energy for a 2D Wigner crystal of electrons in the
plane.\cite{BM:1977} Eq.(\ref{inv_Lap}), restricted to a sphere,
gives\cite{BNT:2000}

\bea \label{Def_Int} H=E_0 &+& \frac{\pi Y}{36} R^2 \sum_{i=1}^{N}
\sum_{j=1}^N q_i q_j
\chi(\theta^i,\psi^i;\theta^j,\psi^j) \nonumber \\
&+& N E_c \ , \eea

\noindent where $E_c$ is a defect core energy, $R$ is the radius
of the sphere and $\chi$ is a function of the geodesic distance
$\beta_{ij}$ between defects with polar coordinates
$(\theta^i,\psi^i$ ; $\theta^j,\psi^j)$: \be \chi(\beta)= 1 +
\int^{\frac{1-\cos\beta}{2}}_0 dz\,\frac{\ln z}{1-z} \ . \ee  The
potential is attractive for opposite charged defects and repulsive
for like-charged defects.  Many predictions of this model are
universal in the sense that they are insensitive to the
microscopic potential. This enables us to make definite
predictions even though the colloidal potential is not precisely
known. It also means that our model system serves as a prototype
for any analogous system with repulsive interactions and spherical
geometry. To further test the validity of this approach, we show a
typical ground state for large M in Fig.~\ref{theoryprediction}.
The system size here is ${\rm R/a}=12$, similar to the droplet in
Fig.~\ref{panel}{\bf D}. The results are remarkably similar to the
experimentally observed configuration in Fig.~\ref{panel}{\bf D};
the only difference is a result of thermal fluctuations, which
break the two defect scars in the experiment. This agreement
between theory and experiment also provides convincing evidence
that these scars are essential components of the equilibrium
crystal structure on a sphere.

The theory predicts that an isolated charge +1 disclination on a
sphere is screened by a string of dislocations of length ${\rm
cos}^{-1}(5/6)R \approx 0.59 R$.\cite{BNT:2000} One can use the
variable linear density of dislocations to compute the total
number of excess dislocations $N$ in a scar. One finds that $N$
grows for large $({\rm R/a})$ as $\frac{\pi}{3}\left[\sqrt{11} -
5\,{\rm cos}^{-1}(5/6)\right]\frac{{\rm R}}{{\rm a}} \approx
0.41\frac{{\rm R}}{\rm a}$, independently of the microscopic
potential. This prediction is universal, and is in remarkable
agreement with the experiment, as shown by the solid line in
Fig.~\ref{NvsR}.

\begin{figure}
\centering
\includegraphics[width=3in]{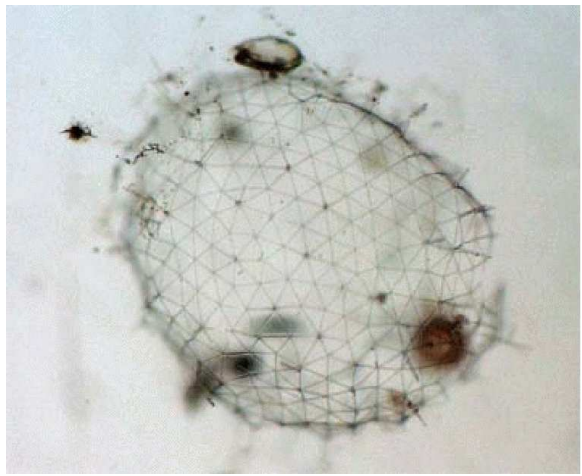}
\caption{The polyhedral siliceous cytoskeleton of the unicellular
ocean organism {\em Aulosphaera}.} \label{aulo}
\end{figure}

\begin{figure}
\centering
\includegraphics[width=2.5in]{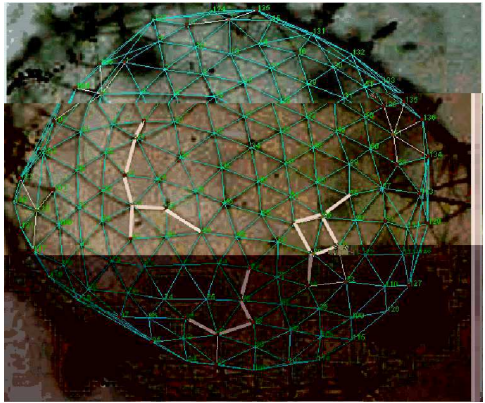}
\caption{The Delaunay triangulation of the {\em Aulosphaera}
cytoskeleon above. } \label{aulotri}
\end{figure}

We expect these scars to be widespread in nature. They should
occur, and hence may be exploited, in sufficiently large stiff
viral protein capsids, giant spherical fullerenes, spherical
bacterial surface layers (s-layers), provided that the spherical
geometry is not too distorted. Terminating strings of heptagons
and pentagons might serve as sites for chemical reactions or even
initiation points for bacterial cell division\cite{Sleytr:2001}
and will surely influence the mechanical properties of spherical
crystalline shells.

The polyhedral siliceous cytoskeleton of the unicellular
non-motile ocean organism {\em Aulosphaera} (a member of the
species {\em Phaeodaria}\cite{ORA:1983}) is shown in
Fig.\ref{aulo}. A triangulation revealing three scars, two of
which are branched, is shown in Fig.\ref{aulotri}. The skeleton
itself is such a perfect triangular lattice that it coincides with
the Delaunay triangulation determined by its vertices. The case of
viral capsids has been analyzed in Ref.\cite{LMN:2003}, where it
is shown that, rather than scarring, icosahedral packings become
unstable to faceting for sufficiently large virus size, in analogy
with the buckling instability of disclinations in two-dimensional
crystals.\cite{SN:1988,NP:1987}

%\begin{figure}[ht]
%\centering
%\includegraphics[width=2.5in]{self_ass.eps}
%\caption{Controlled self-assembly.} \label{fig__selfass}
%\end{figure}

Scarred spherical crystals may provide the building blocks (atoms)
of micron-scale molecules\cite{K:2003,N:2002} and materials. While
topology dictates the overall number of scars (12), the details of
the geometry and defect energetics determine the length and
structure of the scars themselves. It is possible that scarred
colloidosomes will ultimately yield complex self-assembled
materials with novel mechanical or optoelectronic
properties.\cite{K:2003}

\begin{figure}
\centering
\includegraphics[width=4in]{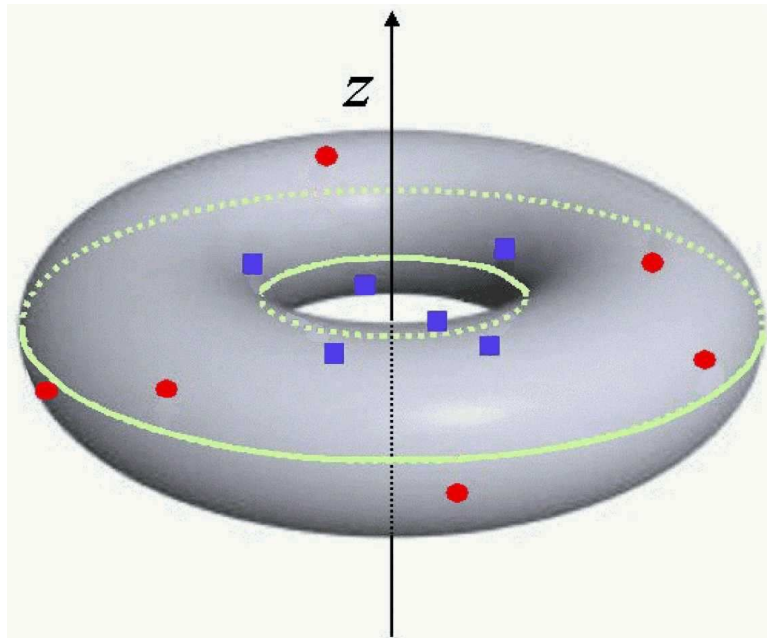}
\caption{A typical ground state for a toroidal hexatic. Five-fold
disclinations are shown as solid circles (red) and 7-fold
disclinations as solid squares (blue).} \label{torus}
\end{figure}

New structures arise if one changes the structure of the colloid
of the topology of the surface they coat. Nelson has analyzed the
case of nematic colloids coating a sphere.\cite{N:2002} In this
case the preferred number of elementary disclination defects is 4,
allowing for the possibility of colloidal atoms with tetrahedral
functionality and sp$^3$-type bonding. The case of toroidal
templates with 6-fold bond-orientational (hexatic) order has been
analyzed recently.\cite{BNT:2003} It is found that defects are
energetically favored in the ground state for fat torii or
moderate vesicle size. A schematic of a ``typical" ground state is
shown in Fig.\ref{torus}.

\section{Acknowlegments}

The work described here has been carried with a large number of
talented collaborators without whom the work would never have been
completed. My thanks go to Alex Travesset, Angelo Cacciuto, Marco
Falcioni, Emmanuel Guitter and Gudmar Thorleifsson.


\begin{thebibliography}{99}
\bibitem{Sphere:12} 12 here is obtained from $6\chi$, where the
Euler characteristic $\chi$ of a space is a topological invariant
equal to 2 for the sphere.
\end{thebibliography}
\end{document}